\documentclass[cameraready]{Interspeech}

\title{mmWave Radar Aware Dual-Conditioned GAN for Speech Reconstruction of Signals With Low SNR
}

\author[affiliation={1}, orcid=0009-0004-5482-2330, equalcontribution, correspondingauthor]{Jash}{Karani}
\author[affiliation={2}, orcid=0009-0008-3705-1668, equalcontribution]{Adithya}{Chittem}
\author[affiliation={2}, orcid=0009-0003-3113-0190, equalcontribution]{Deepan}{Roy}
\author[affiliation={1}, orcid=0000-0001-5620-0472]{Sandeep}{Joshi}

\address{$^1$ Department of Electrical and Electronics Engineering\\
$^2$ Department of Computer Science and Information Systems\\
     Birla Institute of Technology and Science, Pilani, Rajasthan 333031, India
}

\email{\{f20220726, f20220012, f20220004,
sandeep.joshi\}@pilani.bits-pilani.ac.in}


\usepackage{comment}

\begin{document}

\maketitle

\begin{abstract}
    Millimeter-wave (mmWave) radar captures are band-limited and noisy, making for difficult reconstruction of intelligible full-bandwidth speech. In this work, we propose a two-stage speech reconstruction pipeline for mmWave using a Radar-Aware Dual-conditioned Generative Adversarial Network (RAD-GAN), which is capable of performing bandwidth extension on signals with low signal-to-noise ratios (-5 dB to -1 dB), captured through glass walls. We propose an mmWave-tailored Multi-Mel Discriminator (MMD) and a Residual Fusion Gate (RFG) to enhance the generator input to process multiple conditioning channels. The proposed two-stage pipeline involves pre-training the model on synthetically clipped clean speech and fine-tuning on fused mel spectrograms generated by the RFG. 
     We empirically show that the proposed method, trained on a limited dataset, with no pre-trained modules, and no data augmentations, outperformed state-of-the-art approaches for this specific task.
\end{abstract}

\section{Introduction}

The use of millimeter-wave (mmWave) radar for human voice detection has advanced rapidly in recent years. Unlike microphones, radar is contact-free, non-intrusive, and highly directional, making radar particularly effective in complex environments where
microphones often break down \cite{10363766, 9308944, 9931938}. mmWave radar can detect micro-vibrations from a distance, penetrate non-metallic obstructions, and operate passively without direct access to the audio source.
Due to these capabilities, researchers have increasingly explored mmWave radar as a sensing modality for vibration-based speech recovery. However, speech reconstruction from mmWave radar remains intrinsically challenging, as it relies on subtle surface vibrations that are often heavily contaminated by noise and environmental interference. As a result, recovering high-quality and intelligible speech from radar measurements is significantly more difficult than conventional audio enhancement.
Several recent works have attempted to enhance speech from mmWave radar captures with promising results. Nevertheless, many existing approaches rely on large-scale datasets and extensive compute resources, incorporate pre-trained models \cite{han2025hearmmwaveradarendtoend}, evaluate under signal-to-noise ratio (SNR) conditions that may not reflect real-world deployment scenarios \cite{record}, or report metrics that do not reliably correlate with human perceptual quality \cite{waveear}. In this work, we utilize a dataset that is collected using a procedure similar to the one described in \cite{han2025hearmmwaveradarendtoend} under two experimental conditions: (i) direct vibration capture and (ii) vibration capture through an aluminum foil reflector. We show that the proposed pipeline achieves stronger reconstruction quality under constrained resources and adverse low-SNR (-5 dB to -1 dB) settings, addressing a still underexplored and practically relevant problem setting.
The main contributions of this work are as follows.
\begin{itemize}
    \item \textbf{Radar-Aware Dual-conditioned Generative Adversarial Network (RAD-GAN) architecture for mmWave-to-speech reconstruction:} We propose a pipeline to reconstruct intelligible speech through bandwidth extension from extremely low-SNR  (-5 dB to -1 dB), band-limited mmWave Frequency Modulated Continuous Wave (FMCW) radar.

    \item \textbf{Multi-Mel Discriminator (MMD):} We introduce an mmWave radar-based MMD, a two-branch two-dimensional (2D) mel spectrogram discriminator (spectral-norm and weight-norm branches), for stable and realistic reconstruction.
    
    \item \textbf{Two-stage training strategy for stable learning:} We introduce a two-stage training pipeline consisting of pre-training followed by fine-tuning, improving convergence and reconstruction quality.

    \item \textbf{Residual Fusion Gate (RFG) for multi-channel conditioning:} To provide richer conditioning to RAD-GAN, we introduce RFG that produces fused mel spectrograms from noisy inputs and enhanced mel spectrograms produced by a WaveVoiceNet module. 

\end{itemize}
\begin{figure*}[htbp]
    \centering
\includegraphics[width=6.4in, height=2.0in]{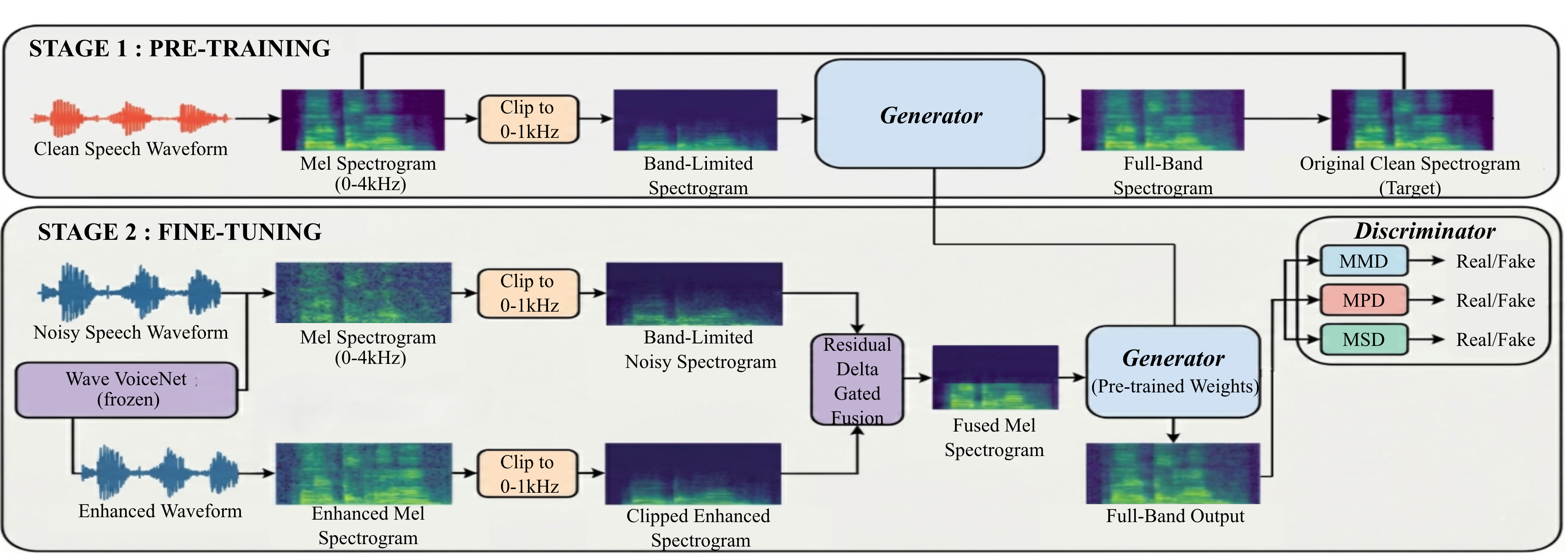}
    \caption{Block diagram of the proposed mmWave radar to speech reconstruction pipeline.}
    \label{fig:flow_diagram}
\end{figure*}




\section{System Model} 

\subsection{RAD-GAN Architecture}
The proposed methodology follows a two-stage pipeline consisting of a pre-training phase on synthetically modified clean speech followed by a fine-tuning phase on real, noisy mmWave-derived recordings, which are combined with the output of the WaveVoiceNet (WVN) to give rich conditioning to the generator as illustrated in Fig.~\ref{fig:flow_diagram}. The complete system is composed of six main components: a HiFi-GAN-based generator, a set of three adversarial discriminators, a WVN module, and a fusion gate used during fine-tuning.

\textbf{Generator}:
We use the HiFi-GAN generator without architectural modification as in \cite{kong2020hifigangenerativeadversarialnetworks}, it maps an 80-bin mel-spectrogram to a waveform through transposed-convolution upsampling with Multi-Receptive Field (MRF) residual fusion blocks, conditioned only on the mel input (no stochastic noise input).
Our contribution lies in the training and conditioning strategy: we first pre-train the generator without discriminators using spectral reconstruction losses on clipped mel inputs. We then perform adversarial fine-tuning with fused mel conditioning to improve perceptual quality while preserving the reconstruction behavior learned during pre-training (Sec.~2.2).

\textbf{Discriminators}:
We use the standard HiFi-GAN waveform discriminators \cite{kong2020hifigangenerativeadversarialnetworks} -- Multi-Period Discriminator (MPD) and Multi-Scale Discriminator (MSD).
By examining the signal at multiple periodic scales, MPD encourages the generator to produce waveforms with consistent rhythmic and harmonic patterns. 
Note that MPD operates on disjoint samples of raw waveforms, whereas MSD operates on smoothed waveforms. MSD encourages the generator to preserve fine local detail while maintaining long-range temporal structure.
In our mmWave radar setting, these discriminators are not sufficient on their own because waveform-level supervision is less reliable under severely degraded phase and limited data. We therefore add a \textbf{Multi-Mel Discriminator (MMD)}, which operates on mel-spectrograms and provides complementary time-frequency signals, improving spectral realism and training stability.
In our implementation, the MMD consists of two parallel 2D convolutional discriminators applied to mel-spectrogram inputs of shape $(B,1,n_{\text{mels}},T)$, where $B$ is the batch size, $n_{\text{mels}}=80$, and $T$ is the number of time frames. The two branches share the same architecture but use different normalization schemes, i.e., one employs spectral normalization for stability, while the other uses weight normalization for flexibility.

Each 2D mel-discriminator processes the mel representation through a stack of convolutional layers ($1{\rightarrow}32{\rightarrow}64{\rightarrow}128{\rightarrow}256{\rightarrow}1$
with kernel $3{\times}3$, padding $1{\times}1$, and strides $(1,2,2,2,1)$) with LeakyReLU activations after each layer. At each layer, intermediate feature maps are stored, capturing hierarchical time–frequency patterns. A final convolution produces a patch level score map rather than a single scalar, enabling the discriminator to assess local realism across different regions of the spectrogram. These scores are flattened and used for adversarial training, while the intermediate feature maps are leveraged for feature-matching loss.
The scores contribute to the standard GAN objectives, while the feature maps are used to align the generator’s internal representations with those of real speech, improving stability and reducing artifacts.

\textbf{WaveVoiceNet Module}:
The WaveVoiceNet model was originally created \cite{waveear} under the premise that large receptive fields are critical for reconstructing speech from mmWave spectrograms. 
We adopt the exact WaveVoiceNet configuration as \cite{waveear} and noticed that it is strong at magnitude-domain transformation but less reliable for phase quality when used as a standalone enhancer. Therefore, we use it as an additional conditioning branch for HiFi-GAN during fine-tuning.

\textbf{RFG}:
We fuse noisy and WVN conditioning using a residual gate, inspired by prior gated/residual formulations \cite{srivastava2015highwaynetworks,he2015deepresiduallearningimage,li2017ideal,YI2021103947}. Let $\mathbf{M}_{n}, \mathbf{M}_{w} \in \mathbb{R}^{B \times F \times T}$ denote noisy-mel and WVN-mel, respectively, then
\begin{equation}
\begin{gathered}
\mathbf{G}=\sigma\mspace{-2mu}\left(\mathrm{Conv}_{1\times1}\left([\mathbf{M}_{n};\ \mathbf{M}_{w}-\mathbf{M}_{n}]\right)\right),\\
\mathbf{M}_{f}=\mathbf{M}_{n}+\sigma(a)\,\mathbf{G}\odot\left(\mathbf{M}_{w}-\mathbf{M}_{n}\right),
\end{gathered}
\end{equation}
where $[\cdot ; \cdot]$ is channel-wise concatenation, $\odot$ is element-wise multiplication, and $a$ is a learnable scalar logit. $\mathbf{M}_n$ is the carry baseline, $(\mathbf{M}_w-\mathbf{M}_n)$ is the WVN residual correction, $\mathbf{M}_f$ is the fused mel, $\mathbf{G}\in[0,1]^{B\times F\times T}$ is a local mask, and $\sigma(a)$ is a global correction scale. The gate uses a pointwise Conv1D ($2F\!\rightarrow\!F$, i.e., $160\!\rightarrow\!80$), operating frame-wise and learning cross-frequency mixing across mel bins without temporal smoothing. This residual form allows fallback toward $\mathbf{M}_n$ when WVN cues are unreliable and amplify WVN cues in reliable regions. We initialize the gate bias and $a$ to $-2.0$ for conservative early training.

\textbf{Loss Functions}:
Let $x$ denote the target clean waveform and $\hat{x}=G(s)$ the generated waveform, where the generator is conditioned on a mel-spectrogram $s$ derived from our radar processing pipeline. Although the input conditioning is band-limited to 0-1kHz during training, all adversarial and feature-based losses are computed over the full 0-4kHz bandwidth of $x$ and $\hat{x}$. This design explicitly penalizes errors made during bandwidth extension, ensuring that the model is not only faithful to the observed low-frequency content but also produces plausible high-frequency structure.



We follow HiFi-GAN~\cite{kong2020hifigangenerativeadversarialnetworks} for the standard least-squares adversarial and feature-matching objectives, $\mathcal{L}_{G}^{\mathrm{adv}}$, $\mathcal{L}_{D}^{\mathrm{adv}}$ and $\mathcal{L}_{G}^{\mathrm{fm}}$ respectively, and use them unchanged in Stage 2 across all sub-discriminators (MPD, MSD, and MMD), summing losses over branches. Full formulations are given in~\cite{kong2020hifigangenerativeadversarialnetworks}.
For WVN pre-training, we use the original MSE reconstruction objective from~\cite{waveear}. We then add task-specific reconstruction losses for the mmWave setting.
Let $\phi(\cdot)$ be the mel-spectrogram transform used for the loss computation.
We use an L1 mel loss with high-frequency weighting, as
%
%
\begin{equation}
\mathcal{L}_{G}^{\mathrm{mel}}
=
\lambda_{\mathrm{mel}} \,
\mathbb{E} \Bigg[
\frac{1}{n_{\text{mels}} \, T}
\sum_{m,t} w_m
\left| \phi(x)_{m,t} - \phi(\hat{x})_{m,t} \right|
\Bigg],
\end{equation}
where $n_{\text{mels}} = 80$, $T$ is the total number of time frames, $\lambda_{\mathrm{mel}} = 45.0$, $w_m = 5.0$ for mel bins above a cutoff $f_c$ and $w_m = 1$ otherwise, and $\mathbb{E}[\cdot]$ denotes the expectation operator.  
This weighting places greater emphasis on reconstruction errors in the upper bands, which is critical for successful bandwidth extension.
To further stabilize training and improve spectral fidelity, we add a multi-resolution short-time Fourier transform (MR-STFT) loss \cite{SteinmetzauralossAL}, which is given as
\begin{align}
\mathcal{L}_{G}^{\mathrm{mrstft}} = \lambda_{\mathrm{stft}}\,
\mathrm{MRSTFT}(x,\hat{x}),
\end{align}
where $\lambda_{\mathrm{stft}} = 5.0$.
%
In Stage 1, we deliberately exclude all adversarial and feature-matching losses and train the generator using only spectral reconstruction objectives as
\begin{align}
\mathcal{L}_{G}^{\mathrm{pre}}
= \mathcal{L}_{G}^{\mathrm{mel}} + \mathcal{L}_{G}^{\mathrm{mrstft}}.
\end{align}
This choice isolates the learning of bandwidth extension and stable waveform synthesis from the instability of GAN training. Once the generator has learned a reliable low-to-high frequency mapping, adversarial supervision is introduced in Stage 2 to refine perceptual quality.
During fine-tuning, we optimize the losses as
\begin{align}
\mathcal{L}_{G}
&= \lambda_{\mathrm{adv}}
\left(
\sum_{D \in \mathcal{D}} \mathcal{L}_{G,D}^{\mathrm{adv}}\right)
 + \mathcal{L}_{G}^{\mathrm{fm}}
 + \mathcal{L}_{G}^{\mathrm{mel}}
 + \mathcal{L}_{G}^{\mathrm{mrstft}}, \\ \textnormal{and} \,\,\,\,
\mathcal{L}_{D}
&= \sum_{D \in \mathcal{D}} \mathcal{L}_{D}^{\mathrm{adv}}.
\end{align}
Here, $\lambda_{\mathrm{adv}}=0.5$ controls the relative contribution of the summed adversarial loss during fine-tuning. We use MSD, MPD, and the proposed MMD in $\mathcal{D}$, so both waveform and mel-level realism are enforced simultaneously. Detailed code can be found at \url{https://github.com/chitadi/RADGAN}.

\subsection{Experimental Setup}
\textbf{Dataset}:
The dataset used was provided as part of the RASE 2026 Challenge \cite{rase2026_dataset}.
The dataset includes paired radar-captured (noisy) and microphone-recorded (clean) speech. TI AWR2243BOOST mmWave FMCW radar was used to capture radar data through a glass wall. Each clean audio has radar-captured data obtained in two scenarios:

Task 1: Direct diaphragm vibration. The radar captures speaker diaphragm vibrations directly. Task 1 contains 6,093 paired speech samples (5,334 train, 759 validation).

Task 2: Secondary surface vibration. The radar captures vibrations from an aluminum foil placed near the speaker diaphragm. Task 2 contains 5,978 paired speech samples (5,229 train, 749 validation).

All speech samples are sampled at 8 kHz and have an average duration of 6.4s. The total duration of the complete paired dataset is $\approx 42$h of speech.
For training consistency, all audio samples were clipped to 4s segments.
The global SNR for radar-captured speech lies in the range of -5\,dB to -1\,dB. Task 2 has a worse segmental SNR as compared to Task 1. 
Our dataset provides synchronized radar vibration data, unlike large publicly available speech datasets \cite{7178964, Yamagishi2019VCTK, Chung_2018}. 
Compared to these datasets, our dataset is significantly smaller. 
Despite limited data, our model generalizes well across both tasks, showcasing its ability to work in realistic and diverse radar-based sensing scenarios.
\textbf{Pre-training}:
We band-limit clean speech to 1\,kHz to match the spectral constraints of mmWave radar. Empirical analysis showed that the most reliable spectral energy lies below 1\,kHz (Task~1: $\approx75\%$, Task~2: $\approx50\%$), while higher bands are mostly noise-dominated. Focusing on low-band cues promotes high-frequency reconstruction instead of memorizing noisy high-band artifacts.
Since, Tasks 1 and 2 share identical clean references, pre-training uses a single clean set of audios: 5,334 clips of 4s ($\sim$5.9h).
Feature extraction is performed using the STFT with parameters $n_{\text{fft}}=1024$, hop size $=128$, and window size $=512$, using a Hann window. The conditioning input to the generator consists of 80-dimensional Mel spectrograms computed with $f_{\min}=0$ Hz and $f_{\max}=1000$ Hz.
The generator is optimized using an MR-STFT loss implemented via the \texttt{auraloss} library \cite{SteinmetzauralossAL}. 
MR-STFT uses three resolutions with FFT sizes $\{256,512,1024\}$ (say $N$), hop sizes $N/4$ ($\{64,128,256\}$), window lengths $N$, and spectral weights $(w_{\mathrm{sc}}, w_{\mathrm{logmag}}, w_{\mathrm{linmag}})=(1.0,1.0,0.0)$.
Training is conducted using the AdamW optimizer with $\beta=(0.9, 0.99)$ and an initial learning rate of $10^{-4}$. An exponential learning rate scheduler with $\gamma=0.999$ is applied at the end of each epoch. Pre-training is run for 66K steps with a batch size of 16 on an NVIDIA A6000 GPU with a total runtime of $\approx$6h. 

\textbf{Fine-tuning}:
In the fine-tuning stage, the model is adapted to real radar-derived noisy inputs while retaining the bandwidth extension capabilities learned during pre-training. Much like the pre-training stage, clipping is applied to the noisy radar signals; moreover, segments are randomly cropped across epochs to increase robustness to temporal variability.
The same STFT parameters and Mel spectrogram configuration are used as in pre-training. Dynamic-range compression is applied to the Mel features using a logarithmic transform.
Before dual-model fine-tuning, WVN is trained separately for 30 epochs using Adam with a learning rate of $10^{-3}$ (batch size 8, gradient accumulation 8), which is used as the conditioning model.
We initialize the RAD-GAN generator from pre-training. During fine-tuning (Fig.~\ref{fig:flow_diagram}), the noisy radar waveform is processed along two paths: WVN produces an auxiliary waveform converted to $\mathbf{m}_{\text{wvn}}$, while the noisy input is converted to $\mathbf{m}_{\text{noisy}}$. We then apply gated residual fusion to generate the fused mel spectrogram which is fed to the RAD-GAN generator.
Fine-tuning then uses the full Stage-2 objective: pre-training reconstruction losses are retained, and adversarial + feature-matching losses from MPD, MSD, and MMD are added.
\begin{figure*}
    \centering
    \includegraphics[width=\linewidth, height = 1.8in]{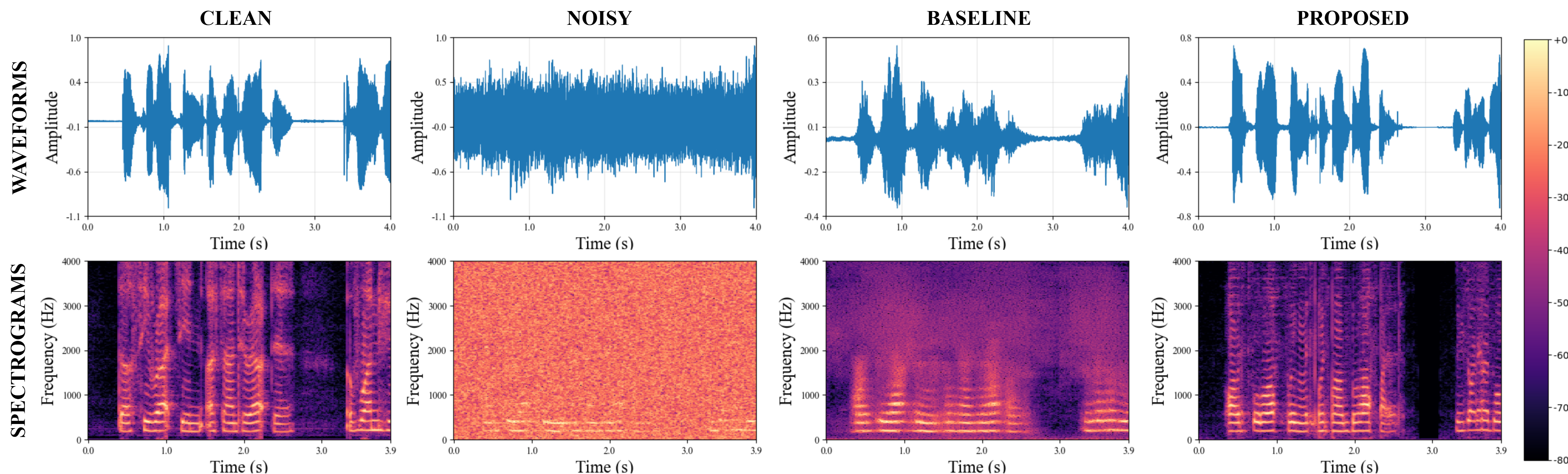}
    \caption{Qualitative comparison on Task 2. Columns from left to right are: clean reference, noisy input, WVN output, and RAD-GAN output. The top row shows the time-domain waveforms, and the bottom row shows the corresponding spectrograms.}
    \label{fig:final_output}
\end{figure*}
Optimization again uses AdamW with $\beta=(0.9,0.99)$ and an initial learning rate of $10^{-4}$. The same exponential learning rate scheduler ($\gamma=0.999$) is applied once per epoch. 
Overall, the proposed model has $87,000,536$ trainable parameters and was fine-tuned for 100K steps with a batch size of 16 for $\approx$14h on an NVIDIA A6000 GPU. 

\section{Results and Analysis}
\textbf{Metrics used}:
We have evaluated the quality of our enhanced speech signal against clean speech signals using the following metrics:
Perceptual Evaluation of Speech Quality (PESQ), which gives a high correlation with subjective opinion and gives robust predictions for a very wide range of conditions \cite{941023, AICHA2012517}. 
Extended Short-Time Objective Intelligibility (ESTOI) \cite{7539284} is a widely used metric that evaluates speech intelligibility. 
Deep Noise Suppression Mean Opinion Score (DNSMOS) \cite{reddy2022dnsmosp835nonintrusiveperceptual} is a neural predictor for human mean opinion score (MOS), which is a subjective measure of human perceived speech quality. 
 Mel-Frequency Cepstral Coefficients Cosine Similarity \cite{9955539,li-etal-2024-cm} calculates the similarity between the MFCC (spectral) features of the clean and enhanced signals.

\textbf{Model comparison}:
To summarize performance across both tasks, we define
\begin{equation}
\begin{aligned}
\widetilde{\mathrm{PESQ}}_i = \frac{\mathrm{PESQ}_i - 1}{4.5 - 1},
\,\,\widetilde{\mathrm{DNSMOS}}_i = \frac{\mathrm{DNSMOS}_i - 1}{5 - 1},
\end{aligned}
\end{equation}
\begin{equation}
Task_i = \frac{\widetilde{\mathrm{PESQ}}_i + \widetilde{\mathrm{DNSMOS}}_i + \mathrm{CSMFCC}_i + \mathrm{ESTOI}_i}{4},
\end{equation}
\begin{equation}
\mathrm{Weighted\ score}
=
0.4\,\mathrm{Task\ 1}
+
0.6\,\mathrm{Task\ 2}.
\end{equation}
We place a higher weight on Task 2 because it is the more challenging setting, so models that remain strong on Task 2 are rewarded accordingly. We use WVN as the main baseline, since it is the most established prior system for mmWave-to-speech transformation \cite{waveear}.
RAD-GAN achieves the best overall \textbf{weighted score (0.333)} and the best \textbf{per-task scores (Task 1: 0.387, Task 2: 0.297)}, outperforming both WVN (0.260) and HiFi-GAN (0.288) as shown in Table 1. Rather than peaking on a single metric, it remains consistently strong across PESQ, ESTOI, CSMFCC, and DNSMOS, indicating a better balance between reconstruction fidelity and perceptual naturalness.
\begin{table}[t]
  \caption{Model comparison results.
    M0: WaveVoiceNet \cite{waveear};
    M1: HiFi-GAN \cite{kong2020hifigangenerativeadversarialnetworks};
    M2: DCCTN \cite{dcctn};
    M3: AP-BWE \cite{10806888};
    M4: DiffWave \cite{kong2021diffwaveversatilediffusionmodel};
    M5: CDiffuSE \cite{lu2022conditionaldiffusionprobabilisticmodel};
    M6: RAD-GAN;
    T1: Task 1;
    T2: Task 2;
    W: Weighted score}
    \label{tab:model_comparison}
  \centering
  \setlength{\tabcolsep}{2.6pt}
  \renewcommand{\arraystretch}{0.8}
  \begin{tabular}{lccccccc}
    \toprule
    & \multicolumn{4}{c}{\textbf{Objective metrics}} & \multicolumn{3}{c}{\textbf{ Scores}} \\
    \cmidrule(lr){2-5}\cmidrule(lr){6-8}
    \textbf{Tag} & \textbf{PESQ} & \textbf{ESTOI} & \textbf{CS} & \textbf{DNS} & \textbf{T1} & \textbf{T2} & \textbf{W} \\
    \midrule
    M0 & $1.302$ & $0.173$ & $0.675$ & $1.558$ & $0.309$ & $0.228$ & $0.260$ \\
    M1 & $1.311$ & $0.144$ & $0.627$ & $2.286$ & $0.332$ & $0.258$ & $0.288$ \\
    M2 & $\textbf{1.547}$ & $0.080$ & $0.377$ & $1.318$ & $0.179$ & $0.167$ & $0.172$ \\
    M3 & $1.174$ & $0.065$ & $0.449$ & $1.472$ & $0.196$ & $0.144$ & $0.165$ \\
    M4 & $1.23$ & $ 0.058$ & $0.288$ & $1.083$ & $0.117$ & $0.100$ & $0.106$ \\
    M5 & $1.175$ & $0.091$ & $0.301$ & $1.225$ & $0.149$ & $ 0.100$ & $0.119$ \\
    M6 & $1.310$ & $\textbf{0.190}$ & $\textbf{0.669}$ & $\textbf{2.688}$ & $\textbf{0.387}$ & $\textbf{0.297}$ & $\textbf{0.333}$ \\

    \bottomrule
  \end{tabular}
\end{table}
 RAD-GAN’s gated residual fusion and two-stage training make better use of low-band cues for stable high-frequency recovery, which is effective in this low-data, low-SNR regime. Notably, these gains are achieved without data augmentation, external pre-trained models, or an explicit phase branch, suggesting that implicit phase recovery via vocoder conditioning is well-suited to low-data, high-noise settings. We additionally show waveform plots to verify temporal fidelity. This is especially important for Task 2, where the input is extremely noisy and provides very limited usable information.
Fig. \ref{fig:final_output} shows the quantitative results: compared with WVN, RAD-GAN reconstructs clearer upper-band harmonics, preserves the 2.6--3.2\,s silence region with less leakage, and follows the clean waveform envelope more closely with sharper onsets/offsets and stronger peak capture. Together, the waveform and spectrogram views show improved spectral structure and temporal realism. Audio examples of RAD-GAN are available online at \url{https://rad-gan-demo-site.vercel.app/}.

\textbf{Ablation Study}:
We conduct an incremental ablation starting from the original HiFi-GAN \cite{kong2020hifigangenerativeadversarialnetworks}. We then add (i) MMD + MR-STFT losses without pre-training, (ii) the pre-training pipeline, and finally (iii) WaveVoiceNet conditioning.
\begin{table}[t]
  \caption{Ablation study results.
    B0: original HiFi-GAN;
    B1: B0 + MMD + MR-STFT;
    B2: B1 + pre-training;
    B3: B2 + WVN conditioning;
    T1: Task 1;
    T2: Task 2;
    W: Weighted Score.
    }
  \label{tab:ablation_results}
  \centering
  \setlength{\tabcolsep}{2.6pt}
  \renewcommand{\arraystretch}{0.8}
  \begin{tabular}{lccccccc}
    \toprule
    & \multicolumn{4}{c}{\textbf{Objective metrics}} 
    & \multicolumn{3}{c}{\textbf{Scores}} \\
    \cmidrule(lr){2-5}\cmidrule(lr){6-8}
    \textbf{Tag} & \textbf{PESQ} & \textbf{ESTOI} & \textbf{CS} & \textbf{DNS} & \textbf{T1} & \textbf{T2} & \textbf{W} \\
    \midrule
    B0 & $\textbf{1.311}$ & $0.144$ & $0.627$ & $2.286$ & $0.332$ & $0.258$ & $0.288$ \\
    B1 & $1.307$ & $0.160$ & $0.588$ & $2.449$ & $0.347$ & $0.251$ & $0.290$ \\
    B2 & $1.286$ & $0.179$ & $0.621$ & $2.639$ & $0.376$ & $0.269$ & $0.312$ \\
    B3  & $1.310$ & $\textbf{0.190}$ & $\textbf{0.669}$ & $\textbf{2.688}$ & $\textbf{0.387}$ & $\textbf{0.297}$ & $\textbf{0.333}$ \\
    \bottomrule
  \end{tabular}
\end{table}
Table~\ref{tab:ablation_results} shows that the overall weighted score improves from \textbf{0.288 (baseline) to 0.333 (+0.045)}. Adding MMD and MR-STFT without pre-training gives a small gain (+0.002), pre-training provides a larger jump (+0.022 over the previous setting), and WVN conditioning gives an additional +0.021. While the weighted score improves monotonically, individual metrics are not strictly monotonic, showing that improving perceived quality does not always improve intelligibility at the same time.
\section{Conclusion}
Speech reconstruction from mmWave radar is difficult because the observations are low-SNR, band-limited, and information-poor (effectively a 1\,kHz$\rightarrow$4\,kHz bandwidth extension setting). We addressed this with a two-stage pipeline: pre-training aided with fine-tuning with WVN-guided gated Mel fusion for robust conditioning. The proposed RAD-GAN is superior to the compared baselines in overall performance, achieving the best weighted score (0.333) and the best results on both tasks defined, while also showing stronger qualitative waveform and spectrogram fidelity. Future work will focus on real-time deployment by reporting latency, followed by model compression through distillation for edge inference.
\section{Generative AI Use Disclosure}
Portions of this manuscript were refined using AI-based writing assistants (e.g., ChatGPT) for grammar, clarity, and style. All scientific content, methodology, and analysis were developed and authored by the researchers.

\section{Acknowledgments}
This work was supported by the Anusandhan National Research Foundation (ANRF), Govt. of India, through its Advanced Research Grant-Mathematical Research Impact Centric Support (ARG-MATRICS) under Grant ANRF/ARGM/2025/001129/TS.  
 
\bibliographystyle{IEEEtran}
\bibliography{mybib}

\end{document}